\documentstyle[prl,aps]{revtex} 
\begin{document}             
\draft
\title{On the signature of tensile blobs in the scattering function
of a stretched polymer}
\author{Carlo Pierleoni}
\address{INFM and Dipartimento di Fisica, Universit\`a degli Studi, 
L'Aquila (Italy)}
\author{Gianluigi Arialdi and Jean-Paul Ryckaert}
\address{D\'epartement de Physique, Universit\'e Libre de Bruxelles, 
Brussels 
(Belgium)} 
\date{\today}
\maketitle
\begin{abstract}
We present Monte Carlo data for a linear chain with excluded volume 
subjected to a uniform stretching. Simulation of long chains (up to 6000 
beads) at high stretching allows us to observe the signature of tensile 
blobs as a crossover in the scaling behavior of the chain scattering 
function for wave vectors perpendicular to stretching. 
These results and corresponding ones in the stretching direction allow 
us to verify for the first time Pincus 
prediction on scaling inside blobs. Outside blobs, the scattering 
function is well described by the Debye function 
for a stretched ideal chain.
\end{abstract}


The notion of blobs in polymer physics is a widespread concept 
which indicates a change of fractal dimensionality of the polymer 
coil across a characteristic length $\xi$ (the blob "size") 
intermediate between the radius of gyration of the chain and the 
local Kuhn segment length. 
Thermic blobs in highly dilute solutions and swollen blobs in semi-
dilute solutions have been observed by careful SANS experiments 
\cite{FBC78} focused on the asymptotic regime of the single chain 
structure factor $S(q) \propto q^{-1/\nu}$ where $\nu$ is the 
scaling exponent. On semi-dilute solutions in good solvent, Farnoux 
et al.\cite{FBC78} observed a clear change from the excluded 
volume statistics ($\nu=0.6$) at length scales characteristic of the 
interior of the blob to the ideal chain statistics ($\nu=0.5$) over 
larger length scales where intramolecular excluded volume forces 
are screened by the presence of other chains. The size $\xi \approx 
1/q^*$ of the ``swollen'' blob is naturally fixed by the cross-over 
wave vector $q^*$ separating the two regimes. 

Dilute solutions of polymers under flow and stretched chains are 
other notorious examples in which the blob concept has been 
invoked under the terminology of ``shear'' blobs and ``tensile'' blobs 
respectively. However SANS experiments are more difficult to 
perform in those cases partly because effects are anisotropic and 
$S(q)$ thus depends upon the direction of $q$ with respect to the 
field.

Onuki has conjectured that shear blobs could develop in long linear 
chains of $N$ bonds in dilute solutions subjected to high shear rates 
\cite{On85}. The shear blob size would correspond to the size of a 
polymer segment of length $n$ ($n<<N$) such that the longest 
relaxation time of this segment is of the order of the inverse shear 
rate. The mere existence of such blobs has not yet been tested 
experimentally although the scattering function of polymer 
solutions under shear has already been probed by SANS 
experiments in the power law regime \cite{Li91}. 
Detecting blobs would require measuring the single chain structure 
factor of very long and highly stretched chains, a measurement 
made difficult by a very unfavorable signal to noise ratio at the 
very low concentrations required to get by extrapolating ``infinite 
dilution" results \cite{LMP97}. 

In this letter, we report Monte Carlo calculations on the structure 
factor of very long chains (up to 6000 monomers) stretched at their 
ends by equal but opposite forces ${\bf f}$ which provide direct 
evidence in favor of the tensile blob concept introduced 20 years 
ago by Pincus and de Gennes \cite{P76,Ge79}. 
These authors suggested that when polymers in good solvent 
are stretched by sufficiently high tensile forces, the coil 
behaves like a stretched ideal chain formed of tensile blobs as 
basic units. Inside the blobs, good solvent statistics persists. This 
picture rests on the assumption that the strong elongation of the 
chain in the field direction prevents monomers distant along the 
polymer contour to interact with each other directly.
In this picture the blob size $\xi_T$ depends on the force intensity 
as $\xi_T = k_BT/f$ where $k_B$ is the Boltzmann constant and 
$T$ the temperature.

Macroscopically the onset of tensile blobs above a threshold value
of the reduced force $\eta=R_o/\xi_T$ ($R_o$ is the unperturbed chain 
end-to-end distance) results in a change of the elastic behaviour of 
the chain. For small stretching the elongation $R_f$ of the chain,
defined as the average projection of the end-to-end vector in the 
field direction, follows a linear Hooke's law $R_f=R_o \eta/3$. In the 
same $\eta$ regime, fluctuations of the square end-to-end distance 
parallel and perpendicular to the force remain at the equilibrium 
value. Above a certain value $\eta^*$ one obtains the following 
power laws
\begin{equation}
        R_f=R_o B \eta^{2/3}
\label{eq:elong}
\end{equation}
\begin{equation}
	\delta R^2_\parallel=R_o^2 C_\parallel \eta^{-1/3}
\label{eq:flucpar}
\end{equation}
\begin{equation}
	R^2_\perp=R_o^2 C_\perp \eta^{-1/3}
\label{eq:flucperp}
\end{equation}
where $B$, $C_\parallel$ and $C_\perp$ are model dependent constants
(independent of N and $\eta$) and the exponent values $2/3$
and $-1/3$ arise from the expected linear
dependence of $R_f$, $R^2_\perp$ and $\delta R^2_\parallel$ on 
$N$ at fixed $f$ in this regime. In what follows, we define 
$R^2_\perp$ as the fluctuations in any direction 
perpendicular to stretching.

A direct test of the elastic behavior, as given by  eqs.
(\ref{eq:elong},
\ref{eq:flucperp}) came long ago from the precursory MC simulation of 
Webman {\it et al.}~\cite{WLK81} which observed the threshold to 
be located around $\eta^*=2$. 
More recent MC study~\cite{WSK94} has been focused on the cross-
over region in the attempt to test renormalization group theory 
predictions~\cite{OOF81}.

We concentrate on the high stretching regime and we provide for 
the first time numerical data for the single chain structure factor. 
Chain lengths of many thousands beads allow us to provide the
first experimental test of Pincus scaling \cite{P76}. Data for $q$ 
perpendicular to the force exhibit a well defined crossover 
between ideal and self-avoiding chain statistics in a 
way analogous to the semidilute case studied by Farnoux {\it et 
al.}~\cite{FBC78}. In the direction parallel to the field and for large 
enough q's,  the scattering function follows the usual 
unperturbed chain asymptotic
behavior when plotted versus an effective scattering 
vector taking into account the change in metric due to stretching,
 as predicted by Pincus \cite{P76}. 
In the low $q$ regime, the generalized Debye function for a 
continuous gaussian chain~\cite{BDODCFJ75} very accurately 
explains all the spectacular features of the scattering function (see 
figure 4).

Our Monte-Carlo study is based on a polymer model which is a 
simple necklace of hard spheres connected by rigid bonds taken as 
length unit. The model is characterized by a hard sphere diameter 
of $\sigma=0.65$, a value which ensures a fully developed 
excluded volume statistics~\cite{LF92}. Equilibrium calculations for 
this model lead to the scaling law: ${R_o=1.011~N^{0.60}}$. 
We combined a reptation algorithm with the Configurational Bias 
Monte Carlo (CBMC) method~\cite{FS96} suitably adapted to include 
the external field influence. A new conformation is 
attempted by suppressing a segment of $k$ beads from one end of 
the chain chosen randomly and constructing a new segment of same 
lenght on the opposite side. 
Leaving out details about the evaluation of the Rosenbluth weights 
appearing in the new segment acceptance probability~\cite{FS96}, 
let us just mention here that the set of $p$ trial directions for the 
single bond reconstruction has been extracted from a probability 
$\propto exp(f cos\phi/k_B T)$ where $\phi$ is the angle 
between the trial direction and the external force.
The number of such trial directions has been chosen in a way to 
minimize the CPU time between two independent configurations. 
Two configurations generated during the MC Markov Chain are 
supposed to be independent when they are separated by 
$(N/2k)^2$ successful reptation moves of $k$ beads. Our 
results are based on statistics of $4000$ independent configurations 
for the chains of length $N<6000$ and $ \approx 1200$ 
independent configurations for the $N=6000$ case.
The optimal number of trial directions $p$ is found to increase 
slightly with the chain lenght (typical value is $p=5$).
Chain lengths of $N= 512, 1024, 2048$ and $6000$ bonds were 
studied at various stretching forces in the range $2\le \eta \le 
40.36$.

Figure 1 shows a Kratky plot of the structure factor of a 6000 bead 
stretched chain, namely $q^2 S(q)$ versus $q$ for the scattering 
vector pointing perpendicularly to the stretching direction. For the 
three particular values of $\eta$ considered, the crossover from 
ideal to excluded volume behavior clearly takes place within the 
$q$ range corresponding to the universal regime of $S(q)$ for 
unstretched chains.  The latter is known to cover a range $3/R_g< 
q <\lambda^{-1}$ where $R_g$ is the equilibrium radius of gyration 
and $\lambda$ the Kuhn segment length~\cite{RDP87}. In the 
present case such range cover two decades from $q_{min}\approx 
0.04$ up to $q_{max} \approx 4.0$.
In figure 1, we also report the Debye function in the perpendicular 
direction~\cite{BDODCFJ75}
\begin{equation}
S(q_\perp)=\frac{2 (N+1)}{X_\perp^2} (e^{-X_\perp}-1+X_\perp)
\end{equation}
where $X_\perp=q_\perp^2 R_{\perp}^2/2$ and values of 
$R_{\perp}^2$
are directly obtained in the simulation.
The Debye function is seen to follow very closely the chain 
structure factor up to the crossover region beyond which all three 
experimental curves converge towards a unique power law 
behavior in $q^{1/3}$ (represented in the figure as a straight line) 
which is reminiscent of the equilibrium universality $S(q)\propto 
q^{-5/3}$.
The observed asymptotic cross-over wave vectors $q^+(\eta)$ 
(taken as the intersection value between the excluded volume 
power law and the Debye functions) should be related to the blob 
size $\xi_T$. Indeed one can easily verify that 
$q^+(\eta)=4.5/\xi_T(\eta) \approx 2 \pi/\xi_T(\eta)$. 

An important issue is about the universality of $S({\bf q},{\bf 
f},N)$. In figure 2, we have plotted the form factors $S(q)/(N+1)$ 
for various chain lengths at the same reduced force $\eta=7$ versus 
$qR_0$. In 
each specific direction we get a distinct universal function, which 
suggests an extension of the well known equilibrium universality of 
$S(q)$ towards
\mbox{$S(q_\perp)= (N+1) f_\perp(q_\perp R_0,\eta)$} and 
$S(q_\parallel)= (N+1) f_\parallel(q_\parallel R_0,\eta)$. Instead if 
we compare $S(q)$ data for different chain lengths at the same 
absolute force intensity $f$, all curves are found to merge for 
$q>q^*$ as it should be.

Along the lines of previous calculations, \cite{WLK81,WSK94} we 
have analyzed the end-to-end vector statistics of stretched chains 
and we confirm that, in the high stretching regime, the elongation 
as well as the squared fluctuations follow the universal laws 
eqs.(\ref{eq:elong}, \ref{eq:flucpar}, \ref{eq:flucperp}) in terms of 
$\eta$. 
The model dependent parameters are $B=0.46$, 
$C_{\parallel}=0.31$ and $C_{\perp}=0.46$. 

Figure 3 shows the systematic evolution of $S(q_{\perp})$ and 
$S(q_{\parallel})$ as $\eta$ increases for $N=6000$.  In the 
direction of the force we observe the appearance of a shoulder which 
becomes more and more pronounced as the chain extends. As 
$\eta$ gets even larger, oscillations develop beyond the initial 
decay. The origin of this behavior can easily be understood by 
comparison with the structure factor for a rigid rod of length $L$ 
with $N+1$ scatterers uniformly distributed which is, for $q$ 
parallel to the rod \cite{On85},
\begin{equation}
	S(q)= \frac{2 (N+1)}{(qL)^2} (1- cos(qL))
\label{eq:sqrod}
\end{equation}

In figure 3, the comparison with the rigid rod is made for $\eta 
=40.36$, using $L=R_o B \eta^{2/3}$ (see eq.(\ref{eq:elong})). At 
small $q$ the chain can be described as a rod of fluctuating length,   
a point not totally surprising if we note that for $N=6000$ and 
$\eta=40.36$, the average extension is one sixth 
of the polymer contour length! The rigid rod model becomes 
obviously more and more inaccurate for increasing $q$, that is on 
the length scales where fluctuations start dominating the 
systematic effects. 
Further, we note that beyond $q=10/\xi_T$, indicated by vertical 
bars in the figure, the structure factor remains isotropic. 

Figure 4 shows the structure factor for $q_\parallel$ in the 
$N=6000$ case for four values of $\eta=0.00, 7.00, 13.34, 40.36$. Data for 
$\eta\neq 0$ are also reported versus the effective wave vector 
$\tilde{q}$ defined as
\begin{equation}
	\tilde{q}^2=q^2+\frac{4}{\xi_T^2}~cos^2\theta
\end{equation}
where $\theta$ is the angle between the observation direction and 
the external field~\cite{P76}. 
For $cos\theta\neq 0$, $\tilde{q}$ is a non linear combination of 
the bare $q$ vector and a measure of the systematic stretching effect. For 
$q\gg\xi_T^{-1}$, $\tilde{q}\simeq q$ and we recover the typical 
power law of the equilibrium scattering function. 
In an intermediate region $q^{**} \le q \le 10/\xi_T$ the curve
 bending of 
$S(q_\parallel)$ 
produced by the external force and directly related to the presence 
of a finite systematic stretching, is compensated by plotting 
$S(q_{\parallel})$ versus $\tilde{q}$. This brings back the data on the 
equilibrium power law ~\cite{P76}
\begin{equation}
	S(q_\parallel)\sim \tilde{q}^{-1/\nu}
\end{equation}
Finally in the region $q \le q^{**} $, $\tilde{q}$ becomes almost 
constant so that the data exhibit an apparent divergence when 
plotted against $\tilde{q}$.

As for the perpendicular direction, the continuous gaussian chain 
model can be exploited to explain the $S(q_{\parallel})$ at low $q$.
The generalized Debye function is~\cite{BDODCFJ75}
\begin{equation}
	S(q_\parallel)=2 (N+1) \Re \left(
	{\frac{e^{-X_\parallel}-1+X_\parallel}{X_\parallel^2}}
					\right) 
\label{eq:debpar}
\end{equation}
where $X_\parallel$ is a complex variable defined as
\begin{equation}
	X_\parallel=q_\parallel^2 \frac{R_o^2}{2}~C_\parallel 
                    \eta^{-1/3} + i q_\parallel R_o B 
\eta^{2/3}
\end{equation}
In figure 4 we report the Debye behavior (dashed lines) for the 
three values of $\eta$ and for values of the elongation and the 
parallel fluctuation given by the observed power laws behaviors of
eqs.(\ref{eq:elong},\ref{eq:flucpar}). The agreement between the 
ideal chain prediction and the experimental data is amazingly good 
up to $q$ values close to $q^{**}$.

In conclusion we have simulated long linear chains stretched by a 
uniform force. We have studied the strong stretching regime where 
tensile blobs dominate the elastic behavior of the 
chains. We have performed a careful analysis of the chain 
scattering function and thanks to the large dimension of our chains 
we have been able to directly observe the blob signature as a cross-
over in the universal behavior of the scattering function in the 
direction orthogonal to the force.
We have provided the first ``experimental'' evidence that Pincus 
scaling in terms of an effective wave vector ~\cite{P76} describes 
accurately the entire excluded volume region of the scattering 
function (inside the blobs). At larger length scales the generalized 
Debye function beautifully explains 
the scattering data, in particular at very high extension when the 
chain behaves as a collection of blobs aligned as a linear rod. The 
picture of an ideal chain of $N_b$ blobs of size $\xi_T$ stretched by 
a force $f$ is however only qualitatively supported by our data. 
Such a model requires $R^2_\perp=\delta R^2_\parallel= (1/3) N_b 
\xi_T^2$ together with $R_f~= (1/3) N_b \xi_T^2 (f/k_BT)$ which 
according to the observed behaviour of eqs.(\ref{eq:elong}, 
\ref{eq:flucpar}, \ref{eq:flucperp}) demands $C_\perp= C_\parallel= 
B$. In our results, $C_\perp=B$ is indeed observed but $C_\parallel 
\approx 2/3 C_\perp$.

The relevance of our results with respect to the experimental
situation is 
threefold: 
first, it should be noted that single chain stress-strain 
curves are nowadays directly measured in beautiful experiments 
on single DNA molecules~\cite{BSF92}. However, similar 
experiments on flexible uncharged single polymers in good solvent 
still appear unfeasible. The second and more traditional 
experimental source of information on stretched chains would be 
scattering measurements on highly diluted solvated gels. However, 
well known network heterogeneities of such samples will transmit 
stress on junction points in a uncontroled way, making it very 
difficult to relate observed spectra to single chain statistics. 
Finally, it appears as most likely that the first experimental test of
the features of the chain structure factor discussed in the present
letter will be provided by SANS experiments on sheared polymer
solutions.

We have the pleasure to thank G. Destr\'ee for unvaluable technical 
assistance. We thank P. Lindner, D. Richter and J. Titantah for 
illuminating discussions. We benefited from a generous INFM 
computer grant on the CRAY-T3D parallel computer at Cineca, 
Bologna (Italy).
 

\begin{thebibliography}{10}

\bibitem{FBC78}
B. Farnoux {\it et~al.}, J. Physique {\bf 39},  77  (1978).

\bibitem{On85}
A. Onuki, J. of Phys. Soc. Japan {\bf 54},  3656  (1985).

\bibitem{Li91}
P. Lindner,  in {\em {Neutron, X-ray and Light Scattering}}, edited by P.
  Lindner and T. Zemb (Elsevier science Publishers, 1991).

\bibitem{LMP97}
P. Lindner, L. Mayer, C. Pierleoni, and J.-P. Ryckaert, ILL Report Exp 9-11-383
   (1997).

\bibitem{P76}
P. Pincus, Macromolecules {\bf 9},  386  (1976).

\bibitem{Ge79}
P.-G. de~Gennes, {\em {Scaling Concepts in Polymer Physics}} (Cornell
  University Press, Ithaca,N.Y., 1979).

\bibitem{WLK81}
I. Webman, J. Lebowitz, and M. Kalos, Phys Rev. A {\bf 23},  316  (1981).

\bibitem{WSK94}
M. Wittkop, J.-U. Sommer, S. Kretmeier, and D. Gritz, Phys. Rev. E {\bf 49},
  5472  (1994).

\bibitem{OOF81}
Y. Oono, T. Ohta, and K. Freed, Macromolecules {\bf 14},  880  (1981).

\bibitem{BDODCFJ75}
H. Benoit {\it et~al.}, Macromolecules {\bf 8},  451  (1975).

\bibitem{LF92}
A. Ladd and D. Frenkel, Macromolecules {\bf 25},  3435  (1992).

\bibitem{FS96}
D. Frenkel and B. Smit, {\em {Understanding Molecular Simulation}} (Academic
  Press, San Diego, 1996).

\bibitem{RDP87}
M. Rawiso, R. Duplessix, and C. Picot, Macromolecules {\bf 20},  630  (1987).

\bibitem{BSF92}
S. Smith, L. Finzi, and C. Bustamante, Science {\bf 258},  1122  (1992).

\end{thebibliography}

\begin{figure}
\caption{N=6000. Kratky plot of $S(q)~q^2$ versus $q$ for $\eta=7$ 
(circles), $13.34$ (squares), $40.36$ (crosses). Dashed curves are 
the predictions of the Debye function and the straight line 
represent the excluded volume behavior $S(q)~q^2 \sim q^{1/3}$.}
\label{fig:kratkyperp}
\end{figure}

\begin{figure}
\caption{Universal behavior of $S(q)/(N+1)$ versus $qR_o$ at 
$\eta=7$. Various chain lengths are shown for both $q$ parallel 
(lower curve) and perpendicular (upper curve) to the external field. 
N=512 (circles), 1024 (squares), 2048 (crosses) and 6000 (stars). 
Equilibrium behavior for N=1024  is given by the thick
line.}
\label{fig:universality}
\end{figure}

\begin{figure}
\caption{N=6000. $S(q)$ for $q$ perpendicular (upper curves) and 
parallel (lower curves) to the field for $\eta=7$ (circles), $13.34$ 
(squares), $40.36$ (crosses) versus the bare wave vector $q$. The 
continuous line is the results of the rigid rod model mentioned in 
the text.}
\label{fig:etadependence}
\end{figure}

\begin{figure}
\caption{N=6000. $S(q_\parallel)$ for $\eta=7$ (circles), $13.34$ 
(squares), $40.36$ (crosses) versus the bare wave vector $q$ (lower 
curves) and the rescaled wave vector $\tilde{q}$ (upper curves). 
For each value of $\eta$, the lower boundary $q^{**}(\eta)$ for the 
validity of Pincus' scaling is indicated by a vertical bar. 
Dashed curves are the predictions of the Debye function. Note that 
they are very accurate up to $q^{**}(\eta)$ which is therefore the 
crossover from ideal to excluded volume behavior.}
\label{fig:scalingpar}
\end{figure}

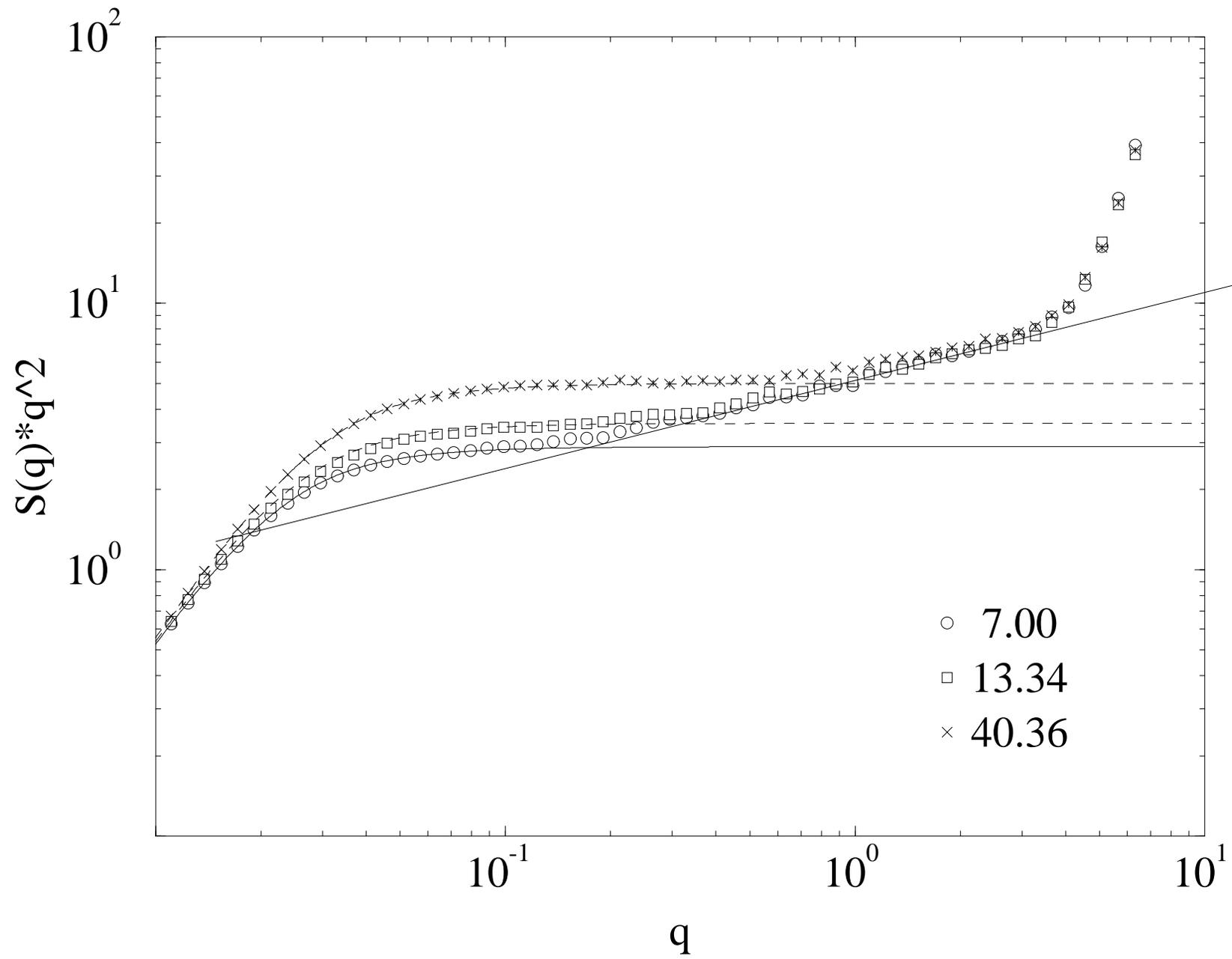
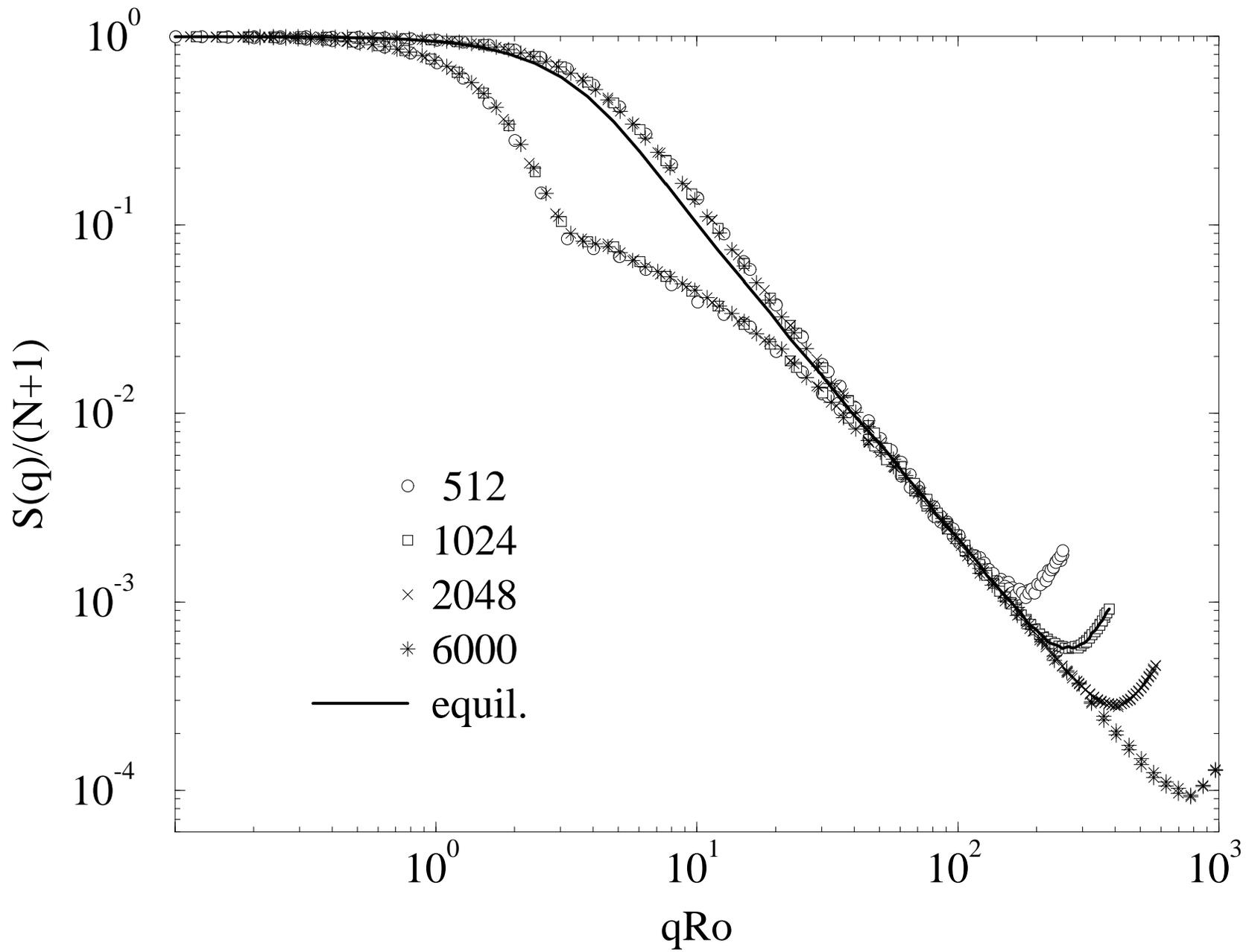
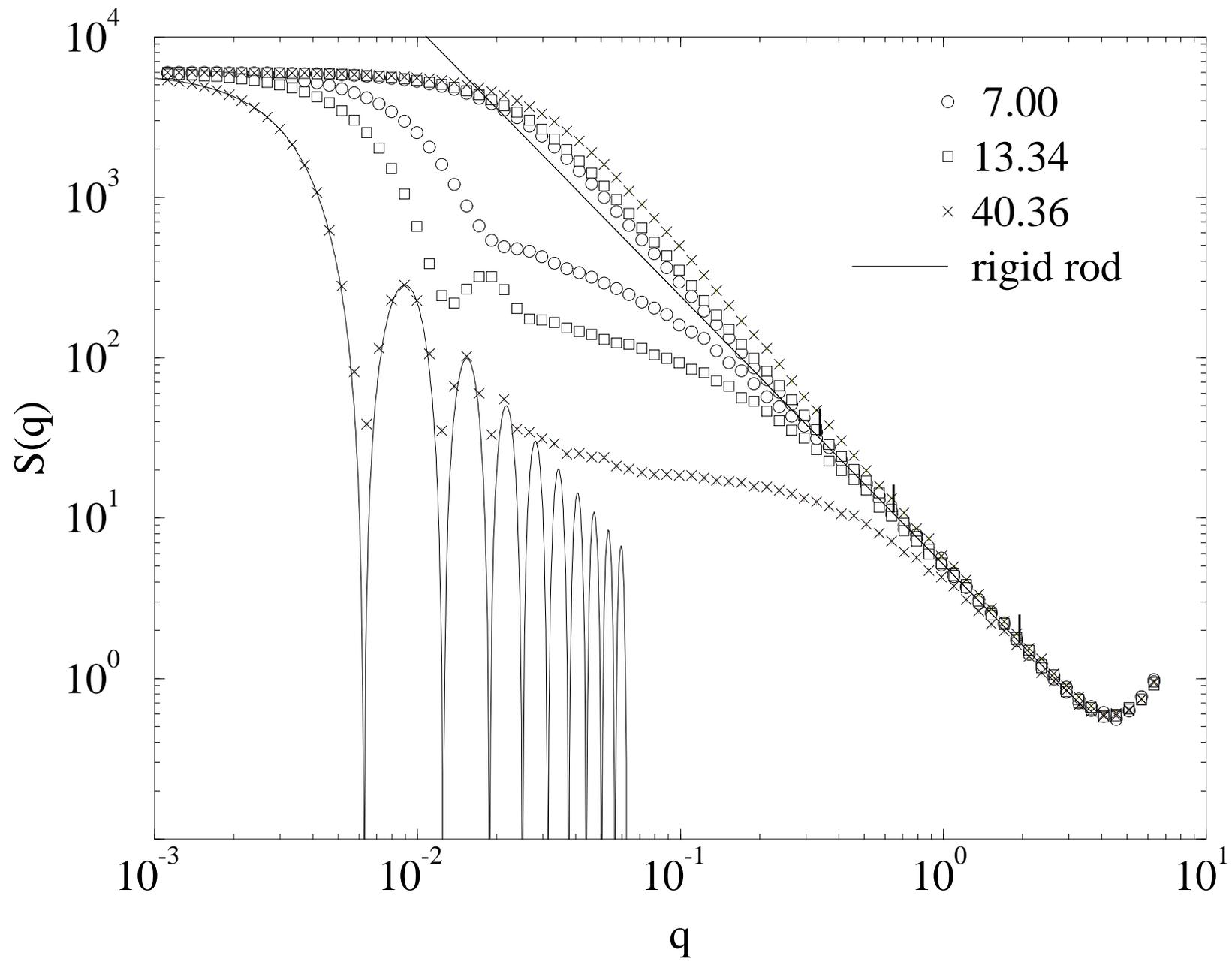
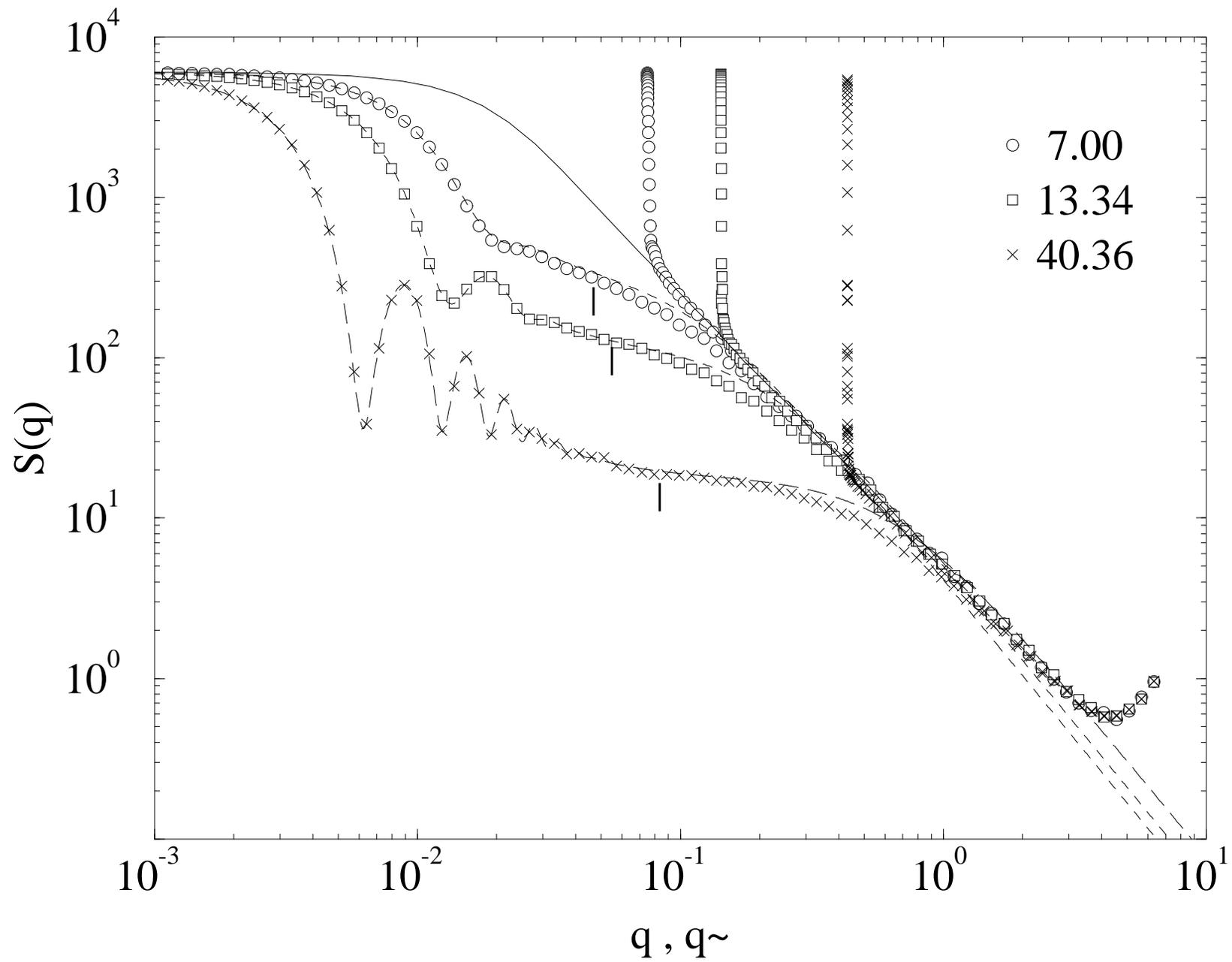
\end{document}